\documentclass[aps,prl,twocolumn,superscriptaddress,showpacs]{revtex4}

\usepackage{amssymb}
\usepackage{amsmath}
\usepackage{graphicx}
\usepackage{color}
\usepackage{textcomp}

\begin{document}

\title{Origin of Ferroelastic Domains in Free-Standing Single Crystal Ferroelectric Films}
\author{I. A. Luk'yanchuk}
\affiliation{Laboratory of Condensed Matter Physics, University of Picardie Jules Verne,
Amiens, 80039, France}
\affiliation{L. D. Landau Institute for Theoretical Physics, Moscow, Russia}
\author{A. Schilling, J. M. Gregg}
\affiliation{Centre for Nanostructured Media, School of Maths and Physics, Queen's University of Belfast, University Road, Belfast BT7 1NN}
\author{G. Catalan, J. F. Scott}
\affiliation{Department of Earth Sciences, University of Cambridge, Downing Street, Cambridge CB2 3EQ}
\date{\today }

\begin{abstract}
The origin of the unusual 90\textdegree  ferroelectric / ferroelastic domains, consistently observed in recent studies on meso and nanoscale free-standing single crystals of BaTiO$_3$ [Schilling et al., Physical Review B, 74, 024115 (2006); Schilling et al., Nano Letters, 7, 3787 (2007)], has been considered. A model has been developed which postulates that the domains form as a response to elastic stress induced by a surface layer which does not undergo the paraelectric-ferroelectric, cubic-tetragonal phase transition. This model was found to accurately account for the changes in domain periodicity as a function of size that had been observed experimentally. The physical origin of the surface layer might readily be associated with patterning damage, seen in experiment; however, when all evidence of physical damage is removed from the BaTiO$_{3}$ surfaces by thermal annealing, the domain configuration remains practically unchanged. This suggests a more intrinsic origin, such as the increased importance of surface tension at small dimensions. The effect of surface tension is also shown to be proportional to the difference in hardness between the surface and the interior of the ferroelectric. The present model for surface tension induced twinning should also be relevant for finely grained or core-shell structured ceramics.

\end{abstract}

 \maketitle

\section*{Introduction}
In 1935 Landau and Lifshitz predicted the appearance of periodic, thermodynamically stable, domains, with oppositely oriented magnetic moments, in ferromagnetic crystals. The existence of these domains minimized the energy of the depolarizing field, caused by the abrupt discontinuity of the spontaneous magnetization at the sample surface \cite{Landau1935}. Additional consideration of the associated domain wall energies allowed them to predict a square root dependence of the domain period on the sample thickness. This square root relationship is often referred to as the Kittel law because of its independent formulation in 1946 by Kittel \cite{Kittel1946}. The law is also valid for 180\textdegree domains in ferroelectric crystals \cite{Mitsui1953, Streiffer2002}, where the unfavourable depolarizing electric field is provided by the abrupt polar discontinuity at the surface. The Landau-Lifshitz-Kittel theory was later extended by Roytburd \cite{Roytburd1976} to describe the behaviour of ferroelastic domains formed as a result of substrate clamping effects in thin film systems.

In previous studies we have shown that the Kittel law works perfectly for a wide class of ferroic materials (ferromagnetic, ferroelectric or ferroelastic) over six orders of magnitude in film thickness \cite{Scott2006, Schilling2006}, and that it can be intuitively expressed in terms of the domain wall thickness \cite{Catalan2007, Stephanovich2005,Guerville2005}. Moreover, we have demonstrated that the Kittel approach can be extended to three dimensional structures \cite{Schilling2006b, Catalan2007b, Schilling2007}, to ferroelectric superlattices \cite{Stephanovich2005} and to multiferroic materials \cite{Catalan2008, Daraktchiev2008}. Much of the experimental work has been in association with observations of periodic 90\textdegree ferroelectric-elastic domains that have been consistently observed in free-standing single crystal thin films (see Fig.1a) and nanowires of BaTiO$_3$. In all our experiments, the size of the 90\textdegree domains as a function of size is indeed found to be well described by a Kittel-Roytburd formalism.

At first glance, however, the very existence of such domains in our free-standing samples is quite surprising. Ferroelastic domains normally appear in response to an external stress (such as that imposed by clamping to a rigid substrate, for example) which forces the sample towards shape preservation in the clamped directions. The domain configuration is such that the macroscopic shape difference between the paraphase and the ferrophase is minimized, while the domain size responds to an equilibrium between domain energy and domain wall energy \cite{Roytburd1976}. However, our BaTiO$_3$ lamella are free standing single crystals and therefore free from epitaxially-induced stress, so they have no Roytburd-like interface-induced elastic driving force. The question, then, is the following: if there is no external stress, what causes the appearance of the ferroelastic domains?

In the present work we explain the appearance of the self-organized domain patterns in free-standing nano-samples of BaTiO$_{3}$ by assuming that the driving stress is provided by an encapsulating surface layer. In much of our experimental work, this encapsulation layer could easily be associated with the surface damage caused by focused ion beam milling. However, we have also observed here that 90\textdegree domains persist, with only slightly altered periodicities, even when surface damage has been repaired by thermal annealing; this suggests that the strain effects may in fact arise intrinsically from surface tension, being therefore unavoidable even in nominally "perfect" free-standing ferroelectric nanostructures.

\begin{figure}[!t]
(a)\includegraphics[width=6.5cm]{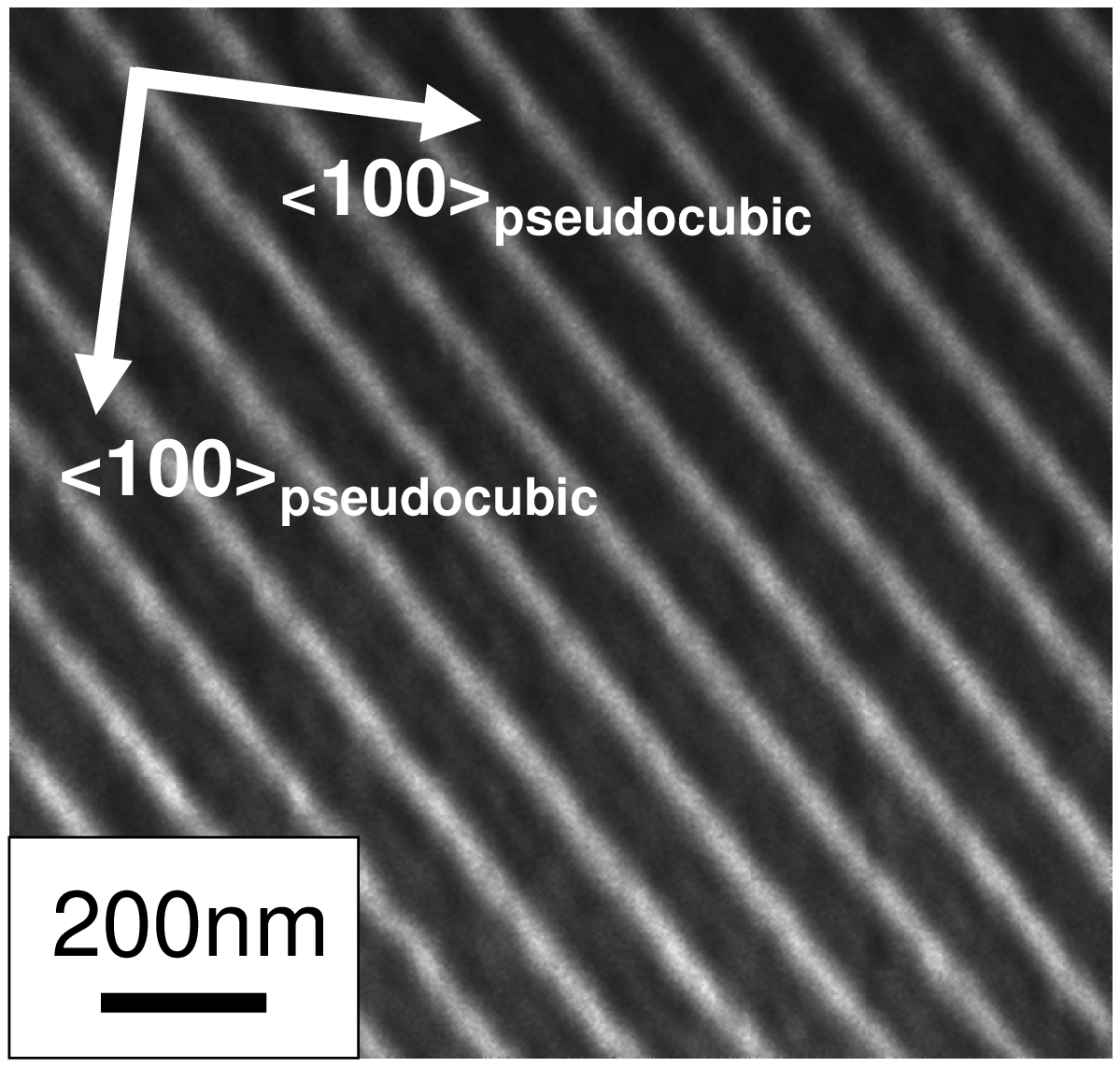}
\vspace{0.5cm}\\
(b)\includegraphics[width=8.0cm]{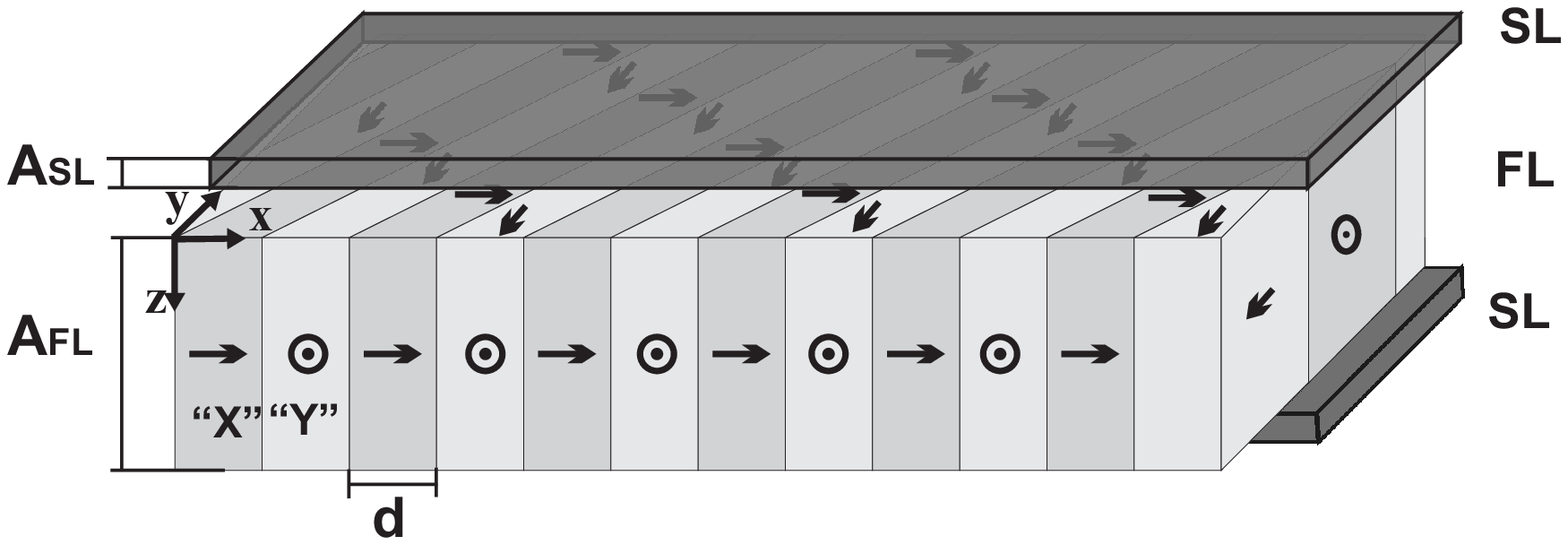} \caption{(Color Online)(a) Scanning
Transmission Electron Microscopy Image of periodic domain structure
in free-standing single crystal lamela of BaTiO$_{3}$. (b) Sketch of our model system, comprising a ferroelectric layer (FL) sandwitched between two surface encapsulation layers (SL) which do not undergo the ferroelastic transformation. The stress imposed by these untransformed "dead" layers onto the ferroelectric layer induces the appearance of 90\textdegree ferroelectric/ferroelastic domains} \label{Plate}
\end{figure}

\section*{Model}
The development of the model focuses on the geometry of the single crystal lamellae first reported in ref.\cite{Schilling2006}, an example of which is shown in Figure 1(a). Figure 1 (b) shows a schematic cross-section of our model representation for such  BaTiO$_{3}$ lamella, with the surface encapsulation layers (SL's) that are potentially responsible for the formation of 90\textdegree domains present and their obeyance to Kittel behaviour. The model comprises a tetragonal ferroelectric/ferroelastic layer (FL) of thickness $A_{FL}$, exemplified by BaTiO$_3$, sandwiched between two cubic paraelectric surface layers (SL) of thickness $A_{SL}$. The surface layers provide the stress for the creation of the ferroelectric-ferroelastic domains in the FL. The assumption of cubic surfaces is supported by the observation that barium titanate nanoparticles have a core-shell  structure with tetragonal interior and cubic surfaces \cite{Takeuchi1994, Tanaka1998}. However, the cubic symmetry is not a key feature of the model: for the SL to impose stress on the interior, it is just sufficient that it does not undergo the phase transition at $T_{C}$. A non-ferroelectric capsule/matrix was also assumed in recent phase-field simulations of domain patterns in ferroelectric nanostructures\cite{Slutsker2008}.

In the present model, the SL thickness is an adjustable parameter to be determined from experimental measurements. The cartesian z-axis is oriented perpendicular to the plane of the lamella, and the x and y axes coincide with the crystal axes of BaTiO$_{3}$ as shown in Fig. \ref{Plate}a. The equilibrium lattice parameters of the FL are assumed to be those of the bulk ferroelectric tetragonal BaTiO$_{3}$ crystal and, in order to minimize depolarization fields, the polarization will tend to lie within the XY plane, pointing parallel to either the X or Y directions. In terms of notation, [$c_{0},a_{0},a_{0}$] if $P$ is parallel to $x$ or [$a_{0},c_{0},a_{0} $] if $P$ is parallel to $y$. The formation of the SL with different equilibrium lattice constants [$b_{0},b_{0},b_{0}$] results in elastic stress, provided by lattice matching at the FL and SL interface.

Consider to start with the simplest possibility when the matching stress is uniform, i.e., there are no stress gradients. Because SL is much thinner than FL, the latter remains undeformed, keeping the equilibrium
BaTiO$_{3}$ lattice parameters ([$c_{0},a_{0},a_{0}$] or [$a_{0},c_{0},a_{0}$]). Deformation concerns only the SL which, because of the matching conditions, should conserve the same XY-plane parameters as the FL but can relax in the Z direction ($s_{zz}^{\prime }=0$). The corresponding deformation energy is caused by the in-plane misfit strains of SL and contains the tension and shear parts \cite{Pompe1993, Speck1994, Speck1995, Pertsev1995} :

\begin{eqnarray}
W_{SL}&=&2\frac{GA_{SL}}{1-\nu }\left( s_{a}^{\prime 2}+2\nu
s_{a}^{\prime }s_{c}^{\prime }+s_{c}^{\prime 2}\right)\notag \\
&=&A_{SL}G\frac{1+\nu }{1-\nu }\left( \delta _{ab}+\delta
_{cb}\right) ^{2}+A_{SL}G\delta _{ca}^{2}  \label{Mono}
\end{eqnarray}%
where the diagonal components of the plain strain tensor of SL
$(s_{a}^{\prime },s_{c}^{\prime },0)=(s_{xx}^{\prime
},s_{yy}^{\prime },s_{zz}^{\prime })$
are expressed via the mismatch parameters:%
\begin{gather}
s_{a}^{\prime }=\delta _{ab}=\frac{a_{0}-b_{0}}{b_{0}},\quad
s_{c}^{\prime }=\delta _{cb}=\frac{c_{0}-b_{0}}{b_{0}},\quad \notag
\\ s_{a}^{\prime }-s_{c}^{\prime }=\delta
_{ca}=\frac{c_{0}-a_{0}}{b_{0}},  \label{mismatch}
\end{gather}%
$G$ is the shear modulus, $\nu$ is the Poisson ratio and the
factor 2 corresponds to two (top and bottom) SLs on both sides of
FL. We assume that $G\approx 40-55\cdot 10^{9}Nm^{-2}$
\cite{Cheng1996} and $\nu \approx 0.28$ \cite{Bradfield1950} are
approximately the same for SL and FL.

Such system can be unstable towards formation of the experimentally observed periodic structure of 90\textdegree ferroelastic domains $X/Y/X/Y$  shown in Fig. \ref{Plate}. Usually, such domains have
45\textdegree domain walls and "head-to-tail" polarization contacts to avoid the formation of depolarization charge. The reason of the instability is that domain formation can reduce the elastic energy (\ref{Mono}) of the SL by allowing the misfit stress to gradually relax inside the FL, within an interfacial region whose thickness is of the order of the domain width $d$. The domains will appear when the elastic energy imposed onto the SL by the ferroelectric is bigger than the energy required for the formation of domain walls. We emphasize that, although the net effect of
ferroelastic twinning is to minimize the macroscopic deformation of the SL, this twinning necessarily requires a strain gradient near
the interface. This is because, while afar away from the interface
the lattice parameters of the FL are those of the bulk
ferroelectric, at the interface they must become closer to cubic in
order to match the SL. Accordingly, an inhomogeneously strained region must appear for which flexoelectric effects may be important (\cite{Cross2006, Catalan2004, Catalan2005, Sharma2009}). The flexoelectric contribution, however, has been left out of this model for the sake of simplicity.

The situation described here is in fact quite similar to the appearance of periodic 90\textdegree domains in epitaxial ferroelectric films strained by
thick undeformable substrates \cite{Pompe1993, Speck1994, Pertsev1995} in which the film/substrate mismatch strain relaxes in the near-surface layer of the film. The only difference is that the
contact elastic SL in our case is thin and its deformation should be considered self-consistently with that of FL. In a way, the model discussed here can be seen as a generalization of the substrate/film models for the case of a substrate that has finite thickness compared with the film.

The mechanism of nonuniform strain relaxation in the
domain-populated FL gives rise to two new energy contributions: the near-surface deformation energy of the ferroelectric layer, $W_{FL}$, and the energy of the domain walls $W_{DW}$. The energy balance between $W_{SL}$, $W_{FL}$ and $W_{DW}$ optimizes the domain period $2d$ and the matching plane lattice constants changing periodically at SL/FL interface from ($c,a$) at X-domain to ($a,c$) at Y-domain.

Before proceeding to the derivation of the total energy of the system
\begin{equation}
W=2W_{FL}+2W_{SL}+W_{DW}  \label{Wtot0}
\end{equation}%
(the factors $2$ correspond to two-side SL and two-side relaxation near-surface layers in FL), we assume that the optimal domain width $d$ is thinner than FL but thicker than SL :%
\begin{equation}
A_{SL}<d<A_{FL}.  \label{ineq}
\end{equation}%
Observed in experiment and discussed in detail later, such hierarchy
simplifies the calculation of the different contributions to
(\ref{Wtot0}).

We express the deformation energy $W_{FL}$ of the ferroelectric layer  in terms of periodically changing strains of domains
$(s_{a},s_{c})/(s_{c},s_{a})$ at the surface of the FL,
\begin{equation}
s_{c}=\frac{c-c_{0}}{c_{0}},\quad s_{a}=\frac{a-a_{0}}{a_{0}},
\label{sasb}
\end{equation}%

These are taken as variational parameters in the
general expression for the elastic energy of the 2-dimensional crystal,
periodically strained as $(s_{a},s_{c})/(s_{c},s_{a})$ and with
relative domain population equal to $\phi $ (See also Eq.(30) in
\cite{Pertsev1995}):
\begin{gather}
W_{FL}=\frac{GA_{FL}}{1-\nu }\left( s_{a}^{2}+2\nu
s_{a}s_{c}+s_{c}^{2}\right) \notag \\ +GA_{FL}(s_{a}-s_{c})^{2}\left[ -2\phi +\frac{2d%
}{A_{FL}}f_{a}\left( \frac{2d}{A_{FL}},\phi \right) \right].
\label{PZ}
\end{gather}%

Adapting Eq.(\ref{PZ}) to the case of equally-populated $(\phi=1/2)$ and thin $(d<A_{FL})$ domains, the universal dimensionless function can be simplified as: $f_{a}\approx A_{FL}/4d+7\zeta (3)/8\pi ^{3}$ \cite{Pertsev1995}, and therefore:

\begin{eqnarray}
W_{FL}=\frac{1}{2}A_{FL}G\frac{1+\nu }{1-\nu }(s_{a}+s_{c})^{2}&+&\frac{1}{2}%
\varkappa ^{-1}dG(s_{a}-s_{c})^{2}, \notag \\ \varkappa &=&{\frac{2\pi ^{3}}{%
7\zeta (3)}}\approx 7.4  \label{PZ1}
\end{eqnarray}%

Physically, the first term in (\ref{PZ1}) corresponds to the average
interface-induced strain that propagates through the whole thickness
$A_{FL}$ of the FL. The second term is produced by the superposition of
alternative strains of an infinite series of domains, compensating
inside the FL and relaxing in the near-surface gradient layer of a thickness of order $d$. Technically this sum is expressed via the zeta-function
$\zeta(3)$, like the electrostatic energy of alternative depolarization
charge compensation in Kittel formula for ferroelectric domains
\cite{Landau1935,Kittel1946}.

Before minimizing (\ref{PZ1}) we can ensure first the vanishing of the largest first (volume) term, selecting $s_a+s_c=0$; this is justified because the domain pattern can compensate for shear strains ($s_a-s_c$) but not for volume changes. The number
of variational parameters then reduces to one: $s\equiv s_{a}=-s_{c}$ and
the second (surface) term takes the form:
\begin{equation}
W_{FL}=2\varkappa ^{-1}dGs^{2}\text{.}  \label{PZ3}
\end{equation}

Consider now the \textit{deformation energy} $W_{SL}$ of the SL
subjected to $2d-$periodic domain-induced surface strains
$(s_{a}^{\prime },s_{c}^{\prime })/(s_{c}^{\prime },s_{a}^{\prime
})$, taking into account that $s_{a}^{\prime }$ and $s_{c}^{\prime
}$ \ are expressed via variational
parameters $s_{a}$ and $s_{c}$ (\ref{sasb}) as:%
\begin{equation}
s_{c}^{\prime }=\frac{c-b_{0}}{b_{0}}\approx s_{c}+\delta
_{cb},\quad s_{a}^{\prime }=\frac{a-b_{0}}{b_{0}}\approx
s_{a}+\delta _{ab} \label{scprim}
\end{equation}

If SL is thinner than the domain structure
period: $(A_{SL}\ll d)$ the periodic surface strain does not manage to
relax across the SL. Then, SL can be presented as a piecewise
$(s_{a}^{\prime },s_{c}^{\prime })/(s_{c}^{\prime },s_{a}^{\prime
})$ strained film, having uniform deformation for each section.
Summing the given by the first part of Eq. (\ref{Mono}) elastic
energies from all the pieces (that are equal because of the
$s_{a}^{\prime }\leftrightarrow s_{c}^{\prime }$ symmetry) and
taking into account the discussed above constrain $s_{a}=-s_{c}=s$ \
we present the elastic energy of SL as a superposition of tension and
shear contributions:
\begin{equation}
W_{SL}=\frac{1}{2}A_{SL}G\frac{1+\nu }{1-\nu }(\delta _{ab}+\delta
_{cb})^{2}+\frac{1}{2}A_{SL}G(2s-\delta _{ac})^{2}.  \label{WSL}
\end{equation}%
Note that the tension contribution:
\begin{equation}
W_{T}=\frac{1}{2}A_{SL}G\frac{1+\nu }{1-\nu }(\delta _{ab}+\delta
_{cb})^{2} \label{WT}
\end{equation}%
does not depend on strain variational parameter $s$ and coincides
with the tension energy of the uniformly deformed SL in
(\ref{Mono}).

Combining (\ref{PZ3}) and (\ref{WSL}) with the energy of domain
walls
\begin{equation}
W_{DW}=\sigma \frac{A_{FL}}{d}  \label{WDW}
\end{equation}%
($\sigma $ is the surface energy density of the domain walls), the
total energy (\ref{Wtot0}) is:
\begin{equation}
W=W_{T}+\sigma \frac{A_{FL}}{d_{0}}\left[
\frac{d}{d_{0}}r^{2}+\varkappa
\frac{A_{SL}}{d_{0}}(r-1)^{2}+\frac{d_{0}}{d}\right] ,  \label{Wtot}
\end{equation}%
where $r=2s/\delta _{ca}$ and
\begin{equation}
d_{0}=\sqrt{\varkappa DA_{FL}}=2.7\sqrt{DA_{FL}}.  \label{param}
\end{equation}%
with the length-scale parameter
\begin{equation}
D=\frac{\sigma }{G\delta _{ca}^{2}}\approx 0.5nm, \label{width}
\end{equation}%
that can be interpreted as "domain wall half-thickness". This value is
ca. one-two lattice constants and agrees with previously reported theoretical and experimental values \cite{Zhirnov1959, Floquet1999, Vanderbilt2002, Shilo2004}.

Minimization of (\ref{Wtot}) over $r$ and $d$ gives the expression for optimal domain width

\begin{equation}
d=\frac{d_{0}}{1-\frac{d_{0}}{\varkappa A_{SL}}}
\label{pseudokittel}
\end{equation}

When $d_{0} \ll \varkappa A_{SL}$, one can neglect the responsible for the upward curvature denominator in (\ref{pseudokittel}) and obtain the Kittel-type dependence (\ref{param}) for $d(A_{FL})$. This
explains why our experimental results could be reasonably well fitted assuming a simple square root dependence of domain periodicity on film thickness \cite{Schilling2006}. Note however that we are still in the thin surface layer limit $A_{SL}<d_{0}$. The opposite (irrelevant for our system) limit $d_{0}<A_{SL}$  will lead to the Roytburd-Pompe-Pertsev situation of thick substrate with domain pattern also obeying the Kittel law but with a different numerical constant. When $d_{0}\rightarrow \varkappa A_{SL}$, the domain width diverges, implying a transition to a mono-domain state. Domains therefore exist only when:
\begin{equation}
d_{0}<\varkappa A_{SL}
\end{equation}%
or, taking into account (\ref{param}) and (\ref{pseudokittel}) when:
\begin{equation}
A_{FL}\lesssim 7.4 \frac{A_{SL}^2}{D}
\end{equation} where D is estimated as (\ref{width}). This represents a rather narrow constraint, providing assumption  (\ref{ineq}) is satisfied.

Substitution of the optimal parameters $r$ and $d$ into (\ref{Wtot})
gives the energy of the domain state
\begin{equation}
W=W_{T}+2\frac{d_{0}}{\varkappa A_{SL}}\left[ 1-\frac{1}{2}\frac{d_{0}}{%
\varkappa A_{SL}}\right] A_{SL} \delta _{ac}^{2}G
\end{equation}%
which is smaller than (\ref{Mono}) (since $2x(1-x/2)\leq 1$ with
$x=d_{0}/\varkappa A_{SL}$) in the limits of applicability of the
theory. This confirms the instability of the ferroelectric
free-standing lamella with surface tension towards domain formation.

\section*{Comparison with experiment}

\begin{figure}[t]
\includegraphics[width=7cm]{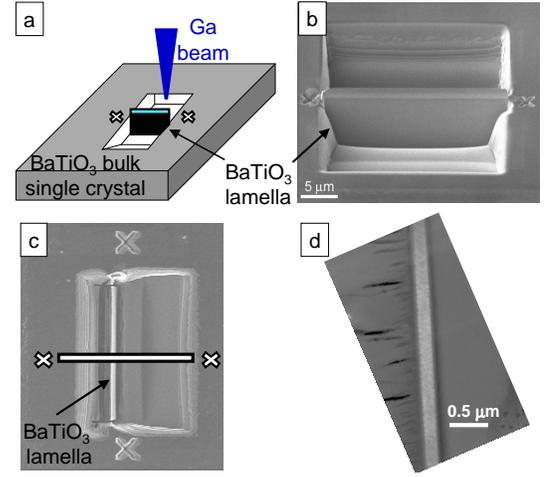}
\caption{\label{FIB}  Different stages of a FIB process to fabricate
a BaTiO$_{3}$ lamella for cross-sectional view: (a) schematic drawing of
a lamella milled by FIB; prior to milling, a protective rim of either Au or Pt (pale blue strip in the top edge of the lamella in the sketch) was deposited in order to help minimize Ga damage during milling, (b) FIB image of a lamella tilted at 45 for
better viewing, (c) drawing of the process to cross-section a
lamella, and (d) cross-sectional TEM image of a lamella milled by
FIB. }
\end{figure}

\begin{figure}[!tb]
\includegraphics[width=8cm]{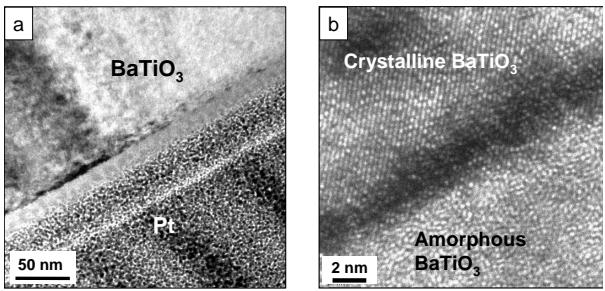}
\caption{\label{DeadLayer} (a) HRTEM image of a cross-sectional BaTiO$_{3}$ lamella after
FIB milling showing a 20nm thick damaged layer at the Pt/BaTiO$_{3}$
interface (the Pt epilayer was deposited to preserve the original surface structure associated with FIB processing of the lamellae) (b) Zoom in on the damage layer, showing its amorphous structure.}
\end{figure}

An attempt was made to use the model developed above to describe the variation in domain periodicity observed for single crystal BaTiO$_{3}$ lamellae  quantitatively. To do this it was noted that the previously published domain period data \cite{Schilling2006} had all been taken from lamellae for which there had been no attempt to repair surface damage caused by focused ion beam (FIB) processing. It was expected that physical 'encapsulation layers' of amorphous BaTiO$_{3}$ should exist on the top and bottom lamellar surfaces. To establish the thickness of the physically damaged layers, cross-sectional transmission electron microscopy was used, with the lamellar cross-sections prepared by FIB according to the schematic shown in figure \ref{FIB}. As can be seen in figure \ref{DeadLayer}, a surface layer of amorphous material, approximately 20nm in thickness, was indeed observed. In conjunction with energy dispersive x-ray data, this layer was categorized as a gallium-impregnated barium titanate glass.

\begin{figure}[!htb]
\includegraphics  [width=8cm]{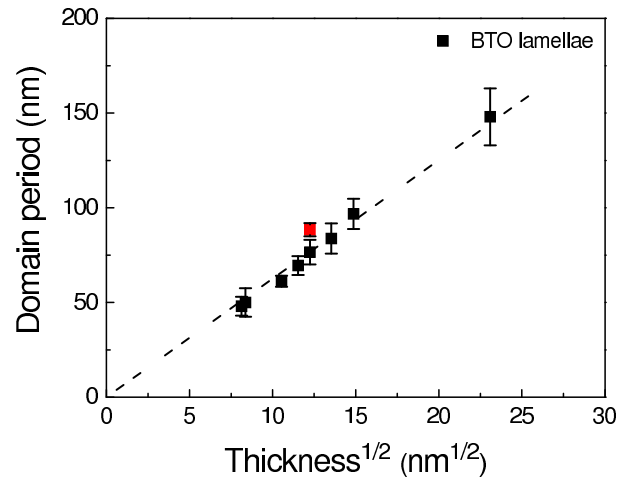}
\caption{\label{fit} (Colour online) Solid data in black colour:
experimentally measured domain width as a function of thickness for
free-standing single crystal films of ferroelectric BaTiO$_{3}$ with
a dead layer thickness of $\sim$ 20 nm. The red colour point
represents the domain width after annealing at 700  \textcelsius for 1 h in
air. For a lamella thickness of 150 nm, the domain width changes
from 76.6 \textpm 6.5 nm (before annealing) to 88.4 \textpm 3.5 nm (after annealing). Dotted line, calculation using eq. (\ref{pseudokittel}) assuming $A_{SL}=20nm$ and a domain wall energy density of $\protect\sigma \approx 3\cdot 10^{-3}Jm^{-1}$ \cite{Zhirnov1959}. The agreement is good without free fitting parameters}
\end{figure}

Substitution of this glassy layer thickness as that of an encapsulation layer ($A_{SL}$) and using a domain wall energy density of $\sigma$ =3\texttimes 10$^{-3}$Jm$^{-1}$ \cite{Zhirnov1959} produced a remarkably good quantitative description of the observed domain periodicity data, as can be seen in figure \ref{fit}. This strong agreement was obtained without any free fitting parameters, and one might naturally conclude that all of the encapsulation suffered by the BaTiO$_{3}$ lamellae was indeed due to the constraint from the ion beam damaged layers.

\begin{figure}[t]
\includegraphics[width=8cm]{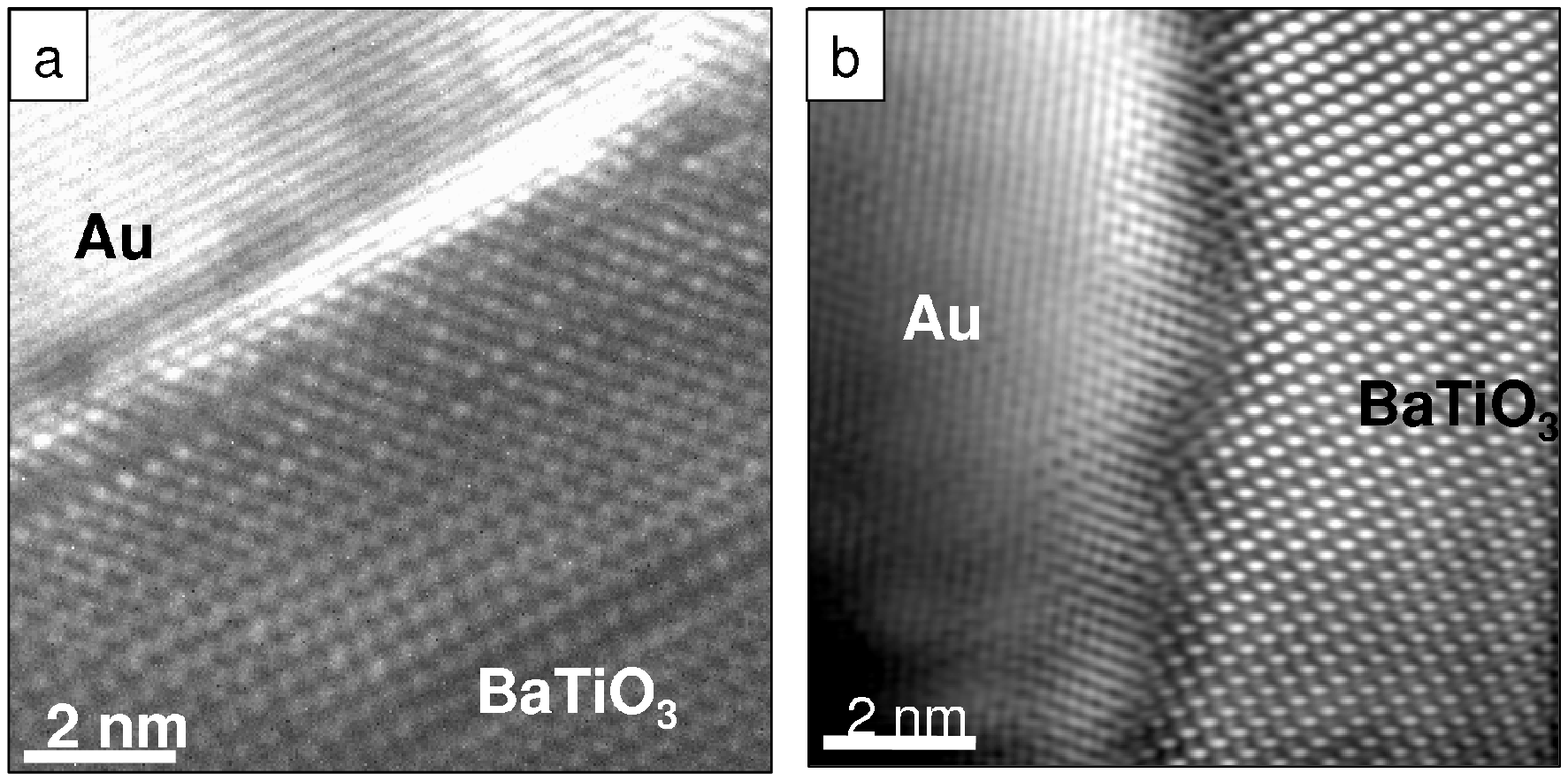}
\vspace{0.3cm}\\
(c)\includegraphics[width=6cm]{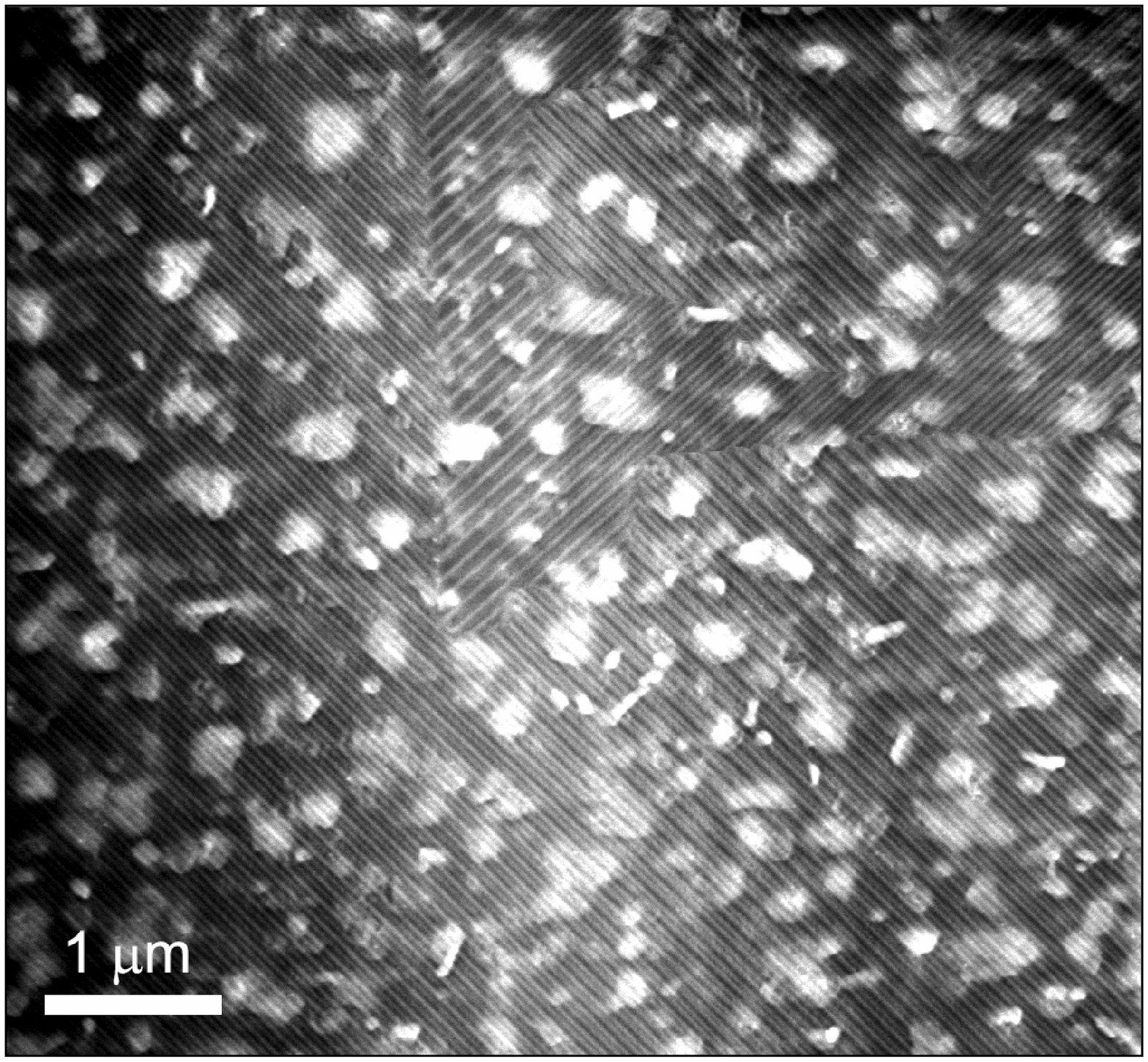}
\caption{\label{Annealed}  (a) HRTEM image of a cross-sectional
BaTiO$_{3}$ lamella after FIB milling and annealing at 700C in air. (b) An inverse FFT image of the Au/BaTiO$_{3}$ interface showing the complete reconstructed structure by post-annealing. Lattice fringes are evident up to the boundary with an Au protective layer, with no evidence of amorphization (c) Plan-view TEM image of the annealed lamella, showing the expelled Ga as clusters on the surface and the persistence of the regular domain pattern.}
\end{figure}

However, we have spent some considerable effort to develop processing methodologies to repair the surface damage caused by focused ion beam processing. Thermal annealing in air at 700 \textcelsius\ has been seen to both recrystallise the damage and expel the implanted gallium (forming thin gallium oxide platelets), recovering pristine single crystal BaTiO$_{3}$ \cite{Saad2004, Schilling2007c, Chang2008}(see figure \ref{Annealed}). If the thermal annealing is performed in oxygen, then functional measurements even suggest that the permittivity of the surface region is the same as that seen in bulk \cite{Saad2004, Chang2008}.

The domain structure seen in a lamella which had been thermally annealed and is expected to have no surface damage is shown in figure \ref{Annealed}-c. The domain appearance is almost identical to the unannealed sample in Figure 1, with the domain walls still of {110}-type, indicating that 90\textdegree domain sets have again formed. While gallium oxide platelets have precipitated on the annealed surface (the white blotches in figure \ref{Annealed}-c), these platelets certainly do not form a continuous layer, and it seems unlikely that they can provide the homogeneous stress needed to induce the 90\textdegree domain sets observed. Further, when the periodicity of the domains is analysed and compared to that obtained in unannealed lamellae, only a slight increase in domain period is observed; the increase in domain period is in fact consistent with an effective increase in the thickness of the lamella by 40nm (that associated with the recrystallisation of the amorphous glassy regions described above). The essential physics at play therefore appears to be unchanged even when the extrinsic damage layer is removed.

Overall, then, it appears that the quantitative agreement between the model and our data may be fortuitous, and that the stress responsible for the domains is not necessarily coming from the glassy barium titanate. Instead, a more fundamental source of stress must be at play for all of the FIBed single crystal barium titanate lamellae investigated to date. This suggests the existence of an intrinsic surface relaxation layer, probably due to surface tension.

The existence of surface layers in BaTiO$_{3}$ has been known for quite some time  \cite{Takeuchi1994, Tanaka1998, Anliker1954, Kanzig1955, Tsai1994, Vanderbilt1997, Vanderbilt1999, Rabe2005, Vanderbilt2007, Rappe2008}, yet there is surprisingly little agreement about their properties. Experimentally, their thickness seems to be on the region of 5-10 nm or more \cite{Takeuchi1994, Tanaka1998, Anliker1954, Kanzig1955, Tsai1994} , whereas first principles calculations give a much smaller value, about 1nm \cite{Vanderbilt1997, Vanderbilt1999, Rabe2005, Vanderbilt2007} . In some works the SL is found to be tetragonal at all temperatures, even above Tc \cite{Anliker1954, Kanzig1955}, where in others it is cubic even below Tc \cite{Takeuchi1994, Tanaka1998}. In fact, the structure of the SL can be rather complex and depends on processing conditions \cite{Rappe2008}. Nevertheless, what is important from the point of view of our model is not so much the exact symmetry of the surface , but the fact that it does not undergo the same ferroelectric/ferroelastic phase transition as the inside of the film. Thus, when the film becomes ferroelectric, it automatically becomes stressed by the untransformed surface layer. The present model requires only the induced stress is isotropic or orthotropic, and this is the case not only with cubic SL's, but also with either amorphous SL or with tetragonal SL provided that the tetragonal axis is out-of-plane.

On the other hand, the elastic energy stored by the surface layer is proportional to its thickness. The outstanding question, then, is whether an intrinsic and very thin SL due to surface tension can lead to the same domain size as would be expected from the thicker encapsulation layers seen in our unnanealed samples.

Indeed, Eq.\ref{pseudokittel} states that the domain size is essentially $d_0$, independent of the SL thickness; however, this is equation is only valid when $d_0 \ll \varkappa A_{SL}$, which is not true if $A_{SL} \approx 1 nm$. On the other hand, our model has implicitly assumed that the stiffness (the shear modulus $G$) is the same for FL and SL, and this is unlikely when the SL is an intrinsic surface-tension layer; for these, the bonds are known to be shorter \cite{Vanderbilt2007} and the SL should be expected to be harder than the FL. It is relatively straight-forward to incorporate the different shear modulus of the surface layer ($G_{SL}$) and the ferroelectric layer ($G_{FL}$) onto the model, by simply substituting $G$ for $G_{FL}$ and $G_{SL}$ in eqs. (\ref{PZ3}) and (\ref{WSL}) respectively. Minimization of the total energy then leads to the new generalized expression for domain size:

 \begin{equation}
d=\frac{d_{0}}{1-\frac{d_{0}}{\varkappa A_{SL}}\frac{G_{FL}}{G_{SL}}}
\label{pseudokittel2}
\end{equation}

This expression is almost identical to (\ref{pseudokittel}) except for the appearance of the factor $G_{FL}\textfractionsolidus G_{SL}$ in the denominator. This factor can compensate for a reduced thickness of the surface layer insofar as its hardness is greater than that of the ferroelectric layer. Presently we have no quantitative estimates for the hardness of the intrinsic epilayer, and we very much encourage the theoretical community to perform first principles calculations of the value of $G_{SL}$. What we can say is that, if the thickness of the SL is only 1-2nm, as suggested by the ab-initio calculations \cite{Vanderbilt1997, Vanderbilt1999, Rabe2005, Vanderbilt2007}, its shear modulus would need to be roughly 10 times bigger than that of the FL to have the same effect on domain size as our 20nm extrinsic encapsulation layer. If, however, the true thickness of the intrinsic SL is ca. 10nm, as suggested by experimental measurements \cite{Takeuchi1994, Tanaka1998, Anliker1954, Kanzig1955, Tsai1994} then the SL need not be more any more rigid than the FL.

\section*{Conclusions}
In sum, we have shown that, even in the absence of rigid substrates or any other source of external stress, ferroelastic twinning can appear due to the self-stress imposed by surface layers. The importance of such layers obviously increases as the size of the system decreases such that this effect becomes particularly important at the nanoscale. Furhermore, there need not be an extrinsic surface layer for the ferroelastic domains to appear; surface tension, which is intrinsic and therefore unavoidable, can also provide the necessary stress for domain formation. Finally, we note that epilayers can be expected to be important not only for isolated nano-ferroelectrics, but also in macroscopic devices such as ceramic capacitors made with nano-powders or core-shell grains.

This work was supported by the EC project FP6-STREP-MULTICERAL and by the French-UK collaboration program  \textquotedblleft Alliance\textquotedblright.

\end{document}